\documentclass[a4paper,fleqn]{cas-sc}

\usepackage[authoryear,longnamesfirst]{natbib}

\begin{document}
\let\WriteBookmarks\relax
\def\floatpagepagefraction{1}
\def\textpagefraction{.001}

\shorttitle{A metabolite glue predicts inverse coupling of AICAR to one-carbon supply}
\shortauthors{}

\title[mode=title]{An allosteric metabolite glue couples AICAR inversely to
one-carbon availability: a theoretical prediction and its therapeutic implications}

\author[1]{Alexei Vazquez}[orcid=0000-0003-2764-3244]
\cormark[1]
\ead{alexei.vazquez@gmail.com}
\affiliation[1]{organization={Nodes \& Links Ltd},
  addressline={Salisbury House, Station Road}, city={Cambridge},
  postcode={CB1 2LA}, country={UK}}
\cortext[1]{Corresponding author}

\begin{abstract}
Formate, an output of mitochondrial one-carbon metabolism, drives a switch from
low to high adenine nucleotide levels in proliferating cells. The kinetic model
of \citet{oizel2020} reproduced this switch under the assumption that de novo
purine synthesis is slaved to biosynthetic demand. Two independent studies have
since identified the physical controller that this assumption stands in for: the
ADP-ribose pyrophosphatase NUDT5 binds and inhibits phosphoribosyl pyrophosphate
amidotransferase (PPAT), the rate-limiting enzyme of purine synthesis, with AMP
acting as an allosteric metabolite glue and the substrate PRPP competing to
dissociate the complex. We replace the demand-slaving assumption with an explicit
PRPP pool and a glue-gated PPAT step, calibrated entirely against the published
binding data. The model makes a sharp, falsifiable prediction: the purine
precursor AICAR is coupled \emph{inversely} to one-carbon availability, rising as
one-carbon units become scarce and clearing when they are abundant. This inverse
coupling is a consequence of the glue: with the feedback removed --- the genetic
state of a NUDT5 knockout --- the model instead predicts AICAR rising with
one-carbon availability, opposite in sign. The prediction reconciles the AICAR
measurements of \citet{oizel2020} without having been fitted to them, and the
model independently reproduces the fall of the PRPP pool reported in NUDT5-deleted
cells. We propose experiments that discriminate the glue from a constitutive
throttle, and discuss the consequences for AMPK signalling and for the
cytotoxicity of thiopurine and antifolate chemotherapeutics, whose action
converges on this same PPAT--NUDT5 node.
\end{abstract}



\maketitle

\section{Introduction}

Proliferating cells consume one-carbon units at a high rate, principally for the
de novo synthesis of purine and thymidine nucleotides, a demand that is under
tight oncogenic control \citep{vidalcruchez2026}. Serine is the dominant
one-carbon donor \citep{yang2016}: its third carbon is oxidised, chiefly in the
mitochondrion, to formate, which is exported to the cytosol, condensed onto
tetrahydrofolate by 10-formyl-tetrahydrofolate synthetase (FTHFS), and used to
formylate the growing purine ring at two positions
\citep{ducker2017,tibbetts2010}. When mitochondrial one-carbon output exceeds the
biosynthetic demand, the excess is released as extracellular formate ---
``formate overflow'' \citep{meiser2018}.

\citet{oizel2020} showed that formate is not merely a substrate but a regulator.
Using a kinetic model together with metabolomics in HAP1 cells, they found that
endogenous or exogenous formate induces a switch from low to high adenine
nucleotide levels, raising the rate of glycolysis and repressing AMP-activated
protein kinase (AMPK). The same coupling operates in leukaemic stem cells, where
inhibiting mitochondrial one-carbon metabolism lowers purine nucleotide levels
and activates AMPK while suppressing mTORC1 \citep{zarou2024}. Their model
coupled formate to energy metabolism through
purine synthesis, but treated purine synthesis itself as slaved to demand: the
rate of one-carbon consumption was set equal to the biosynthetic requirement at
the prevailing growth rate, with no independent enzyme kinetics. This is a
perfect controller with no dynamics --- adequate to reproduce the adenine
nucleotide switch, but silent on the behaviour of the pathway intermediates
between PRPP and the finished nucleotide.

The physical identity of that controller has now been established. Two
independent studies report that the ADP-ribose pyrophosphatase NUDT5 binds
phosphoribosyl pyrophosphate amidotransferase (PPAT), the committed and
rate-limiting enzyme of de novo purine synthesis, and inhibits it.
\citet{witus2026} show that purine nucleotides, and AMP in particular, act as
molecular glues that stabilise the inhibited PPAT--NUDT5 complex, and that the
PPAT substrate PRPP competes at the same site to dissociate it; they term these
endogenous ligands metabolite glues. \citet{strefeler2025}, screening for
regulators of pyrimidine synthesis, independently identify the NUDT5--PPAT
interaction, show that it preserves the PRPP pool, and report that its loss
drives hyperactive purine synthesis at the expense of pyrimidines and confers
resistance to nucleobase analogues. The two studies agree on the mechanism from
different directions, and together they replace the demand-slaving assumption of
\citet{oizel2020} with a measured rate law.

Here we incorporate that rate law into the formate switch model. The purpose is
not to improve the fit to the adenine nucleotide switch --- the glue barely
alters it --- but to ask what the glue predicts for the pathway intermediates,
where the published model was silent. The answer is a single, sharp, falsifiable
statement about AICAR, and we develop its consequences for signalling and for
chemotherapy.

\section{Model}

\subsection{The published model}

The working model of \citet{oizel2020} solves for the free cytosolic
concentrations of AMP, ADP, ATP and formate at steady state, from a formate
balance, a one-carbon balance, an energy balance coupling ATP-consuming growth to
ADP phosphorylation by glycolysis and oxidative phosphorylation, and the
adenylate kinase equilibrium $[\mathrm{AMP}][\mathrm{ATP}]=K[\mathrm{ADP}]^2$.
The proliferation rate follows an effective Michaelis--Menten law in ATP. The
step we replace is the one-carbon balance,
\begin{equation}
f_{\mathrm{CHO}} = \mu\left(2[\mathrm{AMP}]+2[\mathrm{ADP}]+2[\mathrm{ATP}]
+[\mathrm{RNA}]+\tfrac{5}{4}[\mathrm{DNA}]\right),
\label{eq:demand}
\end{equation}
which asserts that one-carbon consumption equals biosynthetic demand. Purine
synthesis has no independent kinetics; it is defined as demand, and the free
adenine pool adjusts to satisfy Eq.~\eqref{eq:demand}. All parameters and their
literature sources are retained from \citet{oizel2020}.

\subsection{A correction to the formate balance}
\label{sec:correction}

The published implementation contains an error in the formate balance, which we
correct here. The third carbon of serine reaches the folate pool by two paths: a
mitochondrial path, in which serine is oxidised to free formate that FTHFS then
condenses onto tetrahydrofolate, and a cytosolic path, in which SHMT1 produces
CH$_2$-tetrahydrofolate directly, with no free formate intermediate. The
cytosolic term therefore belongs in the CHO-THF balance, but in the published
code it was placed in the formate balance, making cytosolic serine catabolism an
artificial source of free formate. This inflates the predicted intracellular
formate and lowers the formate overflow threshold from $42\%$ to $14\%$ of the
FTHFS rate, contrary to the conclusion of \citet{oizel2020} that the
mitochondrial path is the one required for overflow. We place the cytosolic term
in the CHO-THF balance throughout. The correction does not affect the central
prediction of this work (Sec.~\ref{sec:prediction}), which is evaluated with the
cytosolic path inactive.

\subsection{The PPAT--NUDT5 metabolite glue}

\citet{witus2026} report that the octameric PPAT$_4$--NUDT5$_4$ complex assembles
through a nucleotide-independent interface; that PRPP binds the PPAT active site
and dismantles the complex; and that AMP binds the same site, bridges PPAT to
NUDT5, and shields the complex from PRPP-driven disassembly. AMP and PRPP
therefore compete, and we write the inhibited fraction of PPAT as
\begin{equation}
\theta([\mathrm{AMP}],[\mathrm{PRPP}])
= \frac{1+\left([\mathrm{AMP}]/K_a\right)^{n}}
{1+\left([\mathrm{AMP}]/K_a\right)^{n}+[\mathrm{PRPP}]/K_{dP}},
\label{eq:theta}
\end{equation}
which gives a PRPP concentration for half-maximal dissociation of
$\mathrm{IC}_{50}([\mathrm{AMP}])=K_{dP}(1+([\mathrm{AMP}]/K_a)^n)$. The genetic
state of a NUDT5 knockout, in which no glue can form, is recovered by setting the
numerator to zero, i.e.\ $\theta\equiv0$.

We calibrate Eq.~\eqref{eq:theta} entirely against the binding data of
\citet{witus2026}. Their Fig.~3c reports the PRPP IC$_{50}$ for complex
dissociation at three AMP concentrations; the zero-AMP value fixes
$K_{dP}=104~\mu$M, and the shift at $0.1$ and $1$~mM AMP fixes
$K_a=56~\mu$M and $n=1.33$. As an independent check, applied to their
\emph{enzymatic} assay (Fig.~3b, a different modality) Eq.~\eqref{eq:theta}
predicts a half-inhibitory AMP of $332~\mu$M against the measured
$274\pm33~\mu$M --- a $21\%$ error, with no parameter refitted
(Fig.~\ref{fig:cal}a).

\subsection{Extended model}
\label{sec:extended}

We add explicit pools for 10-formyl-tetrahydrofolate (CHO-THF), PRPP, and the two
purine-pathway intermediates that flank the one-carbon-dependent formyl transfer
steps: glycinamide ribonucleotide (GAR, substrate of GART) and
5-aminoimidazole-4-carboxamide ribonucleotide (AICAR, substrate of ATIC). PPAT
initiates the pathway from PRPP under glue inhibition,
\begin{equation}
v_{\mathrm{PPAT}} = P_{\max}\,\frac{[\mathrm{PRPP}]}{K_{\mathrm{PRPP}}+[\mathrm{PRPP}]}
\left(1-\theta\right),
\end{equation}
and the two formyl transfers are limited by one-carbon availability
$c=[\mathrm{CHO}]/(K_c+[\mathrm{CHO}])$,
\begin{equation}
v_{\mathrm{GART}} = g_{\max}\frac{[\mathrm{GAR}]}{K_g+[\mathrm{GAR}]}c,\qquad
v_{\mathrm{ATIC}} = t_{\max}\frac{[\mathrm{AICAR}]}{K_t+[\mathrm{AICAR}]}c.
\end{equation}
The demand-slaving equation~\eqref{eq:demand} is replaced by mass balances on the
new pools and an adenine balance in which purine supply is whatever the pathway
delivers, $\phi\,v_{\mathrm{ATIC}}=\mu(\,[\mathrm{AMP}]+[\mathrm{ADP}]
+[\mathrm{ATP}]+\Pi_{\mathrm{ade}})$, with $\phi=0.625$ the adenine fraction of
purine flux. FTHFS is written reversibly (Appendix~\ref{app:model}), since the
cytosolic synthetase domain of MTHFD1 can run backwards; this matters only for
the cytosolic-serine condition and is otherwise inert. The energy balance, the
adenylate kinase equilibrium and $\mu([\mathrm{ATP}])$ are unchanged. The eight
steady-state balances, all parameters with provenance tags, and the numerical
method are given in Appendix~\ref{app:model}; the full parameter list is
Table~\ref{tab:par}. The model has thirteen added parameters, of which three (the
glue constants) are fitted to \citet{witus2026} and the rest assumed; the
sensitivity analysis of \S\ref{sec:sens} is therefore integral, not optional.

\section{Results}

\subsection{The glue calibration cross-validates and does not reshape the switch}

\begin{figure}[t]
\centering
\includegraphics[width=\linewidth]{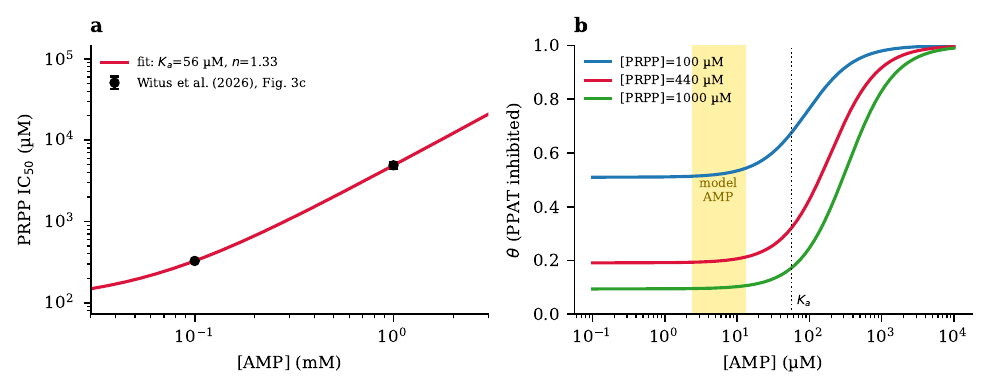}
\caption{Calibration of the PPAT--NUDT5 glue against \citet{witus2026}.
(a)~PRPP IC$_{50}$ versus AMP; the fit to Fig.~3c predicts the independent
enzymatic $K_i$ of Fig.~3b within $21\%$. (b)~Inhibited fraction $\theta$ versus
AMP at three PRPP levels; the gold band is the free AMP range spanned by the
model, lying on the flat foot of the response $4$--$24\times$ below $K_a$.}
\label{fig:cal}
\end{figure}

Constants fitted to the dissociation assay predict the independent enzymatic
$K_i$ within $21\%$ (Fig.~\ref{fig:cal}a), supporting the competitive form of
Eq.~\eqref{eq:theta}. Because AMP binds the same site as PRPP and stabilises the
complex, rather than forming competing binary complexes, the response is
monotonic in AMP with no hook effect; forward and backward parameter sweeps agree
to $<4\times10^{-14}$, so the model shows no hysteresis or bistability.

Across the full range of serine catabolism to formate, the inhibited fraction
$\theta$ moves only from $19.5\%$ to $21.4\%$ (Fig.~\ref{fig:scan}). The reason is
quantitative: the model's free AMP spans $2.3$--$13.1~\mu$M, i.e.\
$0.04$--$0.23\times K_a$, so the AMP-sensing arm of the glue never engages and
$\theta$ is pinned near its PRPP-set floor $\theta\approx1/(1+[\mathrm{PRPP}]/K_{dP})$
(Fig.~\ref{fig:cal}b). The glue is, in this regime, a near-constant throttle
rather than a graded regulator, and the adenine nucleotide switch of
\citet{oizel2020} is essentially unchanged by it. The interest of the glue lies
not in the switch but in the pathway intermediate it controls.

\subsection{Key prediction: AICAR is coupled inversely to one-carbon availability}
\label{sec:prediction}

\begin{figure}[t]
\centering
\includegraphics[width=\linewidth]{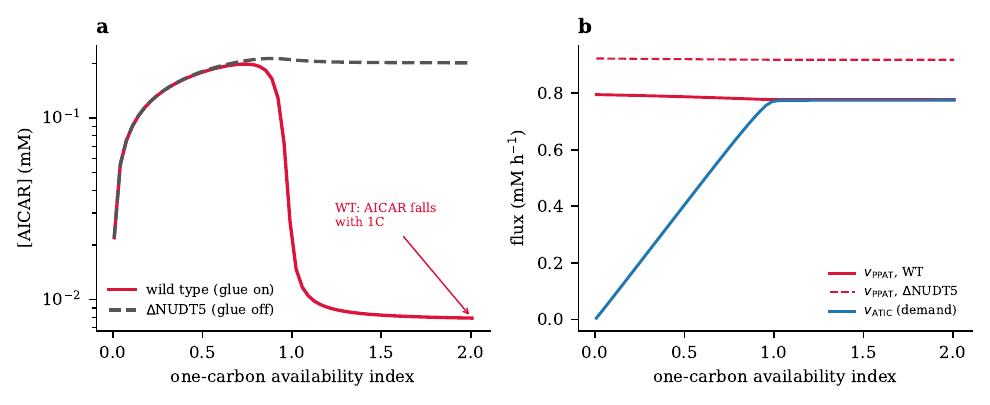}
\caption{The key prediction and its dependence on the glue. (a)~AICAR versus
one-carbon availability in wild type (solid, glue on) and in a NUDT5 knockout
(dashed, glue off). The two curves differ in the \emph{sign} of the coupling.
(b)~PPAT initiation and ATIC completion fluxes; the glue keeps $v_{\mathrm{PPAT}}$
close to demand in wild type, whereas in the knockout PPAT overshoots and AICAR
cannot clear.}
\label{fig:key}
\end{figure}

The extended model predicts that AICAR falls as one-carbon availability rises
(Fig.~\ref{fig:key}a, solid). Over the scan, AICAR rises from $0.022$~mM at the
lowest one-carbon supply to a maximum of $0.198$~mM at intermediate supply, then
collapses to $0.008$~mM when one-carbon units are abundant --- a net inverse
coupling (Spearman $\rho=-0.81$). GAR, the upstream intermediate, falls
monotonically (Fig.~\ref{fig:scan}).

The mechanism is transparent from the balances. At steady state the AICAR pool is
set by the overshoot of PPAT initiation above the rate at which the pathway is
completed,
\begin{equation}
[\mathrm{AICAR}] \approx
\frac{v_{\mathrm{PPAT}}-v_{\mathrm{ATIC}}}{\mu+k_{\mathrm{deg}}},
\label{eq:aicar}
\end{equation}
where $v_{\mathrm{ATIC}}$ is pinned to adenine demand through the adenine balance.
When one-carbon units are scarce, ATIC --- which requires a formyl-THF ---
stalls, PPAT continues to fire from an ample PRPP pool, and AICAR accumulates.
When one-carbon units are abundant, ATIC keeps pace, the glue holds
$v_{\mathrm{PPAT}}$ within $0.002$~mM~h$^{-1}$ of demand, and AICAR clears
(Fig.~\ref{fig:key}b). The non-monotonic peak arises because at the very lowest
one-carbon supply GART also stalls, so flux backs up at GAR before it can reach
AICAR.

This prediction is not an artefact of fitting: no AICAR measurement entered the
calibration. It nonetheless reconciles the metabolomics of \citet{oizel2020},
whose Fig.~3b reports AICAR high in one-carbon-deficient lines (MFT-SHMT1, MFT,
SHMT2) and near-zero in formate-replete and wild-type lines --- a significant
negative association with the one-carbon availability index
($p=5.4\times10^{-4}$) --- and whose Fig.~5d shows, on titrating formate into
MFT-SHMT1 cells, AICAR rising to a maximum near $0.25$~mM formate and then falling
to undetectable levels, the non-monotonic shape of Eq.~\eqref{eq:aicar}.

\subsection{The inverse coupling requires the glue: a NUDT5 knockout should
reverse its sign}
\label{sec:ko}

The inverse coupling is a property of the glue, not of the pathway topology. When
the feedback is removed --- the genetic state of a NUDT5 knockout, $\theta\equiv0$
--- PPAT is no longer restrained, and at high one-carbon availability it overshoots
adenine demand by $0.116$~mM~h$^{-1}$ ($v_{\mathrm{PPAT}}=0.918$ against
$v_{\mathrm{ATIC}}=0.802$) rather than by $0.002$. By Eq.~\eqref{eq:aicar} AICAR
then remains elevated at $0.20$~mM even when one-carbon units are abundant, and
the coupling turns positive (Spearman $\rho=+0.43$; Fig.~\ref{fig:key}a, dashed).
The knockout is thus predicted to \emph{invert the sign} of the AICAR--one-carbon
relationship.

\begin{figure}[t]
\centering
\includegraphics[width=\linewidth]{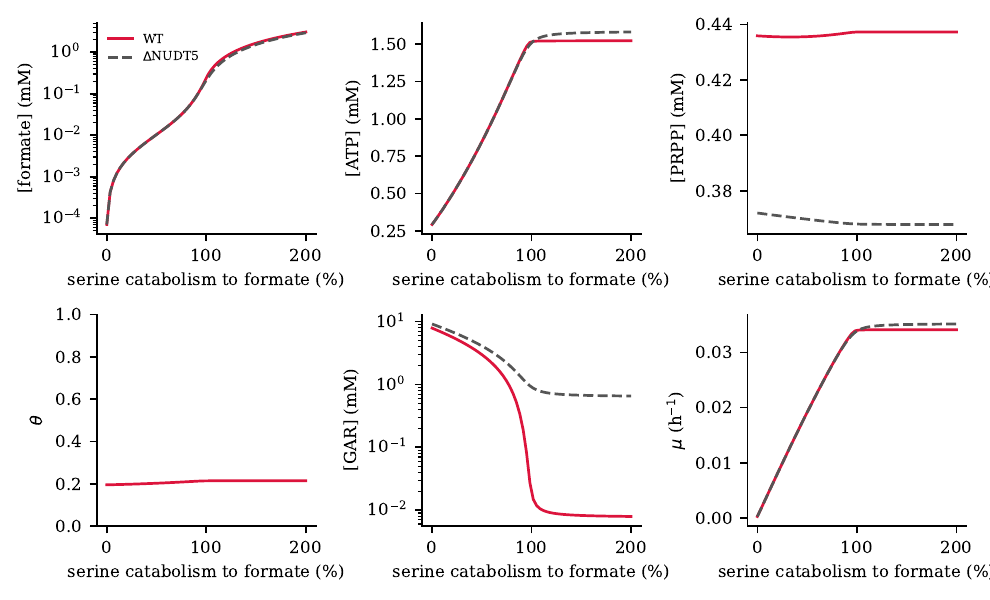}
\caption{Extended model across serine catabolism to formate, in wild type
(solid) and NUDT5 knockout (dashed). The adenine nucleotide switch (ATP) is
essentially unchanged by the glue; PRPP is drawn down in the knockout, matching
\citet{strefeler2025}; $\theta$ is near-constant.}
\label{fig:scan}
\end{figure}

Two independent observations already bracket this prediction. \citet{strefeler2025}
report that NUDT5 deletion depletes the PRPP pool, as PPAT runs unrestrained; our
model reproduces this without adjustment, giving a PRPP pool $16\%$ lower in the
knockout than in wild type (Fig.~\ref{fig:scan}, PRPP panel) because the same
overshoot that elevates AICAR also draws down PRPP. The direction and approximate
magnitude match. What has not been measured is the quantity the model most
sharply predicts: the sign of the AICAR response to one-carbon supply in the
knockout.

\subsection{Magnitude versus mechanism, and the discriminating experiment}
\label{sec:disc}

Because $\theta$ is near-constant, the wild-type AICAR collapse could in
principle be produced by any constitutive restraint on PPAT, not specifically by
the glue. We tested this by removing the glue and instead lowering the PPAT
capacity $P_{\max}$ by a fixed factor (Table~\ref{tab:throttle}). A constitutive
reduction of $\approx18$--$20\%$ reproduces the AICAR collapse as well as the glue
does. The wild-type data therefore constrain the \emph{magnitude} of PPAT
restraint --- enough to bring $v_{\mathrm{PPAT}}$ within $\approx0.03$~mM~h$^{-1}$
of demand --- but not its \emph{mechanism}.

Two considerations nonetheless favour the glue, and one experiment settles it.
First, the required window is narrow ($\geq18\%$) and the glue lands in it
($19.5$--$21.4\%$) without being fitted to any pathway data. Second, and
decisively, the two hypotheses diverge under NUDT5 deletion: removing the glue
raises the overshoot to $0.116$~mM~h$^{-1}$ and inverts the AICAR coupling
(\S\ref{sec:ko}), whereas a cell whose PPAT capacity is constitutively $20\%$
lower, with no glue, continues to clear AICAR normally and shows no inversion.
Measuring the sign of the AICAR--one-carbon relationship in NUDT5-null cells thus
discriminates the glue from a generic throttle. This is the central experimental
proposal of \S\ref{sec:exp}.

\begin{table}[t]
\centering
\caption{A constitutive PPAT throttle mimics the glue in wild type. Scan over
serine catabolism to formate; AICAR$_{\mathrm{hi}}$/AICAR$_{\mathrm{peak}}$ is the
ratio of the high-one-carbon value to the peak. Only the glue additionally
predicts sign inversion under NUDT5 deletion (\S\ref{sec:ko}).}
\label{tab:throttle}
\begin{tabular}{lccc}
\toprule
$P_{\max}$ (\% default) & $v_{\mathrm{PPAT}}$ (high) &
AICAR$_{\mathrm{hi}}$/AICAR$_{\mathrm{peak}}$ & collapses? \\
\midrule
$100$ (glue off) & $0.918$ & $0.95$ & no \\
$85$ & $0.822$ & $0.25$ & no \\
$82$ & $0.805$ & $0.09$ & yes \\
$80$ & $0.787$ & $0.05$ & yes \\
\midrule
glue on ($P_{\max}$ default) & $0.777$ & $0.04$ & yes \\
\bottomrule
\end{tabular}
\end{table}

\subsection{Sensitivity to assumed parameters}
\label{sec:sens}

Each assumed parameter was varied twofold up and down and the scan repeated. Two
conclusions separate by robustness. The qualitative predictions are robust: the
span of $\theta$ across the scan never exceeds $2.2$ percentage points, and AICAR
is elevated at low one-carbon availability in every variation, so the inverse
coupling and its glue-dependence do not depend on the assumed constants. The
quantitative magnitude is not robust: the absolute AICAR concentration spans a
$\sim$640-fold range across the variations, since the steady-state pool is
essentially $v_{\mathrm{PPAT}}/k_{\mathrm{deg}}$ and $k_{\mathrm{deg}}$ is
assumed. The model predicts the \emph{sign and shape} of the AICAR response, not
its absolute level, and the proposed experiments are framed accordingly.

\subsection{The glue is a dual sensor: the AMP arm reports energy charge}
\label{sec:dualsensor}

That $\theta$ is near-constant across the one-carbon scan reflects the regime the
model occupies, not an intrinsic inertness of the feedback. The free AMP that the
glue senses is not a free parameter: it is fixed by the adenylate kinase
equilibrium $[\mathrm{AMP}][\mathrm{ATP}]=K[\mathrm{ADP}]^2$, so it is low when
the energy charge is high. In the proliferating, formate-replete state
($\mathrm{EC}\approx0.95$) free AMP is $2$--$13\,\mu$M, well below $K_a=56\,\mu$M,
and the AMP-sensing arm of Eq.~\eqref{eq:theta} is silent; $\theta$ is set by the
competing PRPP term alone. This is a property of the energy state, not of the
glue.

\begin{figure}[t]
\centering
\includegraphics[width=\linewidth]{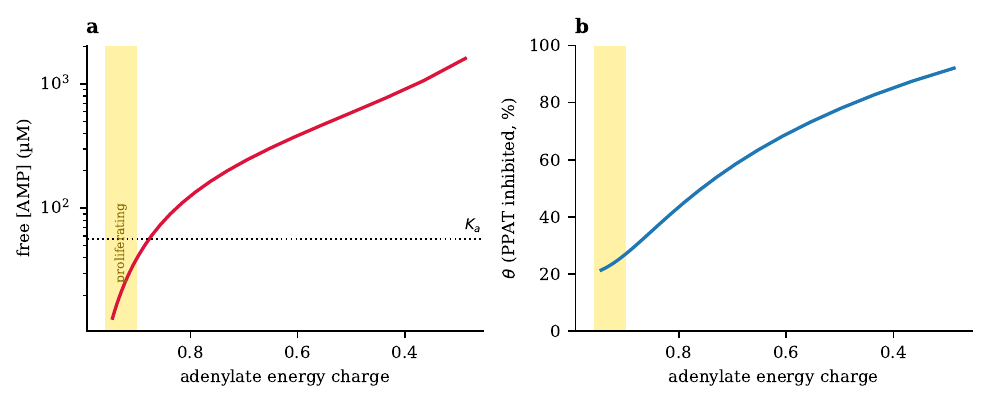}
\caption{The AMP arm of the glue is an energy-charge sensor. Free AMP is fixed by
the adenylate kinase equilibrium and rises steeply as the energy charge falls
(a), crossing the glue affinity $K_a$ and driving the inhibited fraction $\theta$
from its PRPP-set floor towards saturation (b). The gold band marks the
proliferating, high-energy-charge regime in which the AICAR prediction is
evaluated and the AMP arm is silent.}
\label{fig:ec}
\end{figure}

Lowering the energy charge --- by reducing the ATP-supply capacity at fixed
one-carbon supply --- raises free AMP steeply (Fig.~\ref{fig:ec}a). It crosses
$K_a$ near $\mathrm{EC}\approx0.84$ and reaches $222\,\mu$M at
$\mathrm{EC}\approx0.71$, engaging the AMP arm and driving $\theta$ from $21\%$ to
above $50\%$ (Fig.~\ref{fig:ec}b). The glue is therefore a \emph{dual sensor}:
substrate-gated through PRPP at high energy charge --- the regime of the AICAR
prediction above --- and energy-gated through AMP once the energy charge falls.
Because the AMP arm and AMPK read the same rising free AMP, the glue shuts down
the ATP- and PRPP-consuming purine pathway under exactly the conditions that
activate AMPK (\S\ref{sec:ampk}), a coherent energy-conservation response.

This also resolves an apparent tension with measured nucleotide levels. The free
AMP of the model ($2$--$13\,\mu$M) is far below the total AMP reported by
metabolomics ($10^2$--$10^3\,\mu$M), but the two are different quantities: the
glue constants were measured against free AMP, whereas metabolomics reports
total, largely protein-bound AMP. A free AMP as high as $100$--$300\,\mu$M would
require an energy charge near $0.7$ (Fig.~\ref{fig:ec}a), i.e.\ an energy-stressed
rather than a proliferating state.

\section{Proposed experiments}
\label{sec:exp}

The prediction is designed to be falsifiable with established assays. We propose,
in order of decisiveness:

\begin{enumerate}
\item \textbf{Sign of the AICAR--one-carbon relationship, wild type versus
NUDT5-null.} Titrate formate into SHMT2-deficient cells, or a serine/glycine
gradient into wild-type cells, and quantify intracellular AICAR by LC--MS across
the gradient, in parallel in NUDT5-null cells. The model predicts a
\emph{negative} slope in wild type and a \emph{positive} slope in the knockout
(Fig.~\ref{fig:key}a). A sign inversion confirms the glue; an unchanged negative
slope indicates a NUDT5-independent throttle and refutes the specific mechanism.
This is the single experiment that separates the two hypotheses of
\S\ref{sec:disc}.
\item \textbf{Glue-pocket point mutant.} Repeat (1) in cells carrying the
NUDT5 glue-interface mutation (L217A/K218A of \citealp{witus2026}), which abolishes
AMP-dependent inhibition while leaving NUDT5 expressed and its hydrolase activity
intact. The mutant should phenocopy the knockout (positive slope), isolating the
glue from any scaffolding or catalytic role of NUDT5.
\item \textbf{PRPP and AICAR co-measurement.} The model ties the AICAR overshoot
and the PRPP draw-down to the same PPAT over-firing (\S\ref{sec:ko}). Measuring
both in the same samples across the one-carbon gradient tests the predicted
anti-correlation between the AICAR excursion and the PRPP pool, linking our
prediction to the PRPP phenotype of \citet{strefeler2025}.
\item \textbf{AMPK readout.} Since AICAR is an AMP mimetic and AMPK activator
(\S\ref{sec:ampk}), assay phospho-AMPK (Thr172) and phospho-ACC across the
gradient. The model predicts AMPK activity tracks AICAR, i.e.\ falls with rising
one-carbon supply in wild type and fails to fall in the knockout, providing an
orthogonal, antibody-based confirmation independent of metabolite quantification.
\end{enumerate}

The absolute AICAR concentrations are not predicted (\S\ref{sec:sens}); all four
experiments are framed as tests of \emph{sign}, \emph{ordering} and
\emph{monotonicity}, which are robust model outputs.

\section{Discussion}

\subsection{AICAR as a node linking one-carbon supply to energy signalling}
\label{sec:ampk}

The prediction acquires its significance from what AICAR is. Beyond its role as a
purine-pathway intermediate, AICAR (as its monophosphate, ZMP) is an AMP mimetic
that allosterically activates AMPK, the master sensor of cellular energy charge
\citep{corton1995,hardie2012}; the pharmacological AMPK activator AICA-riboside
acts through precisely this species. An inverse coupling of AICAR to one-carbon
availability therefore places one-carbon metabolism upstream of AMPK through a
route independent of the adenylate energy charge itself.

This reframes an observation of \citet{oizel2020}, who reported that formate
represses AMPK and attributed it in part to the fall of AICAR. The present model
supplies the mechanism and, crucially, its controller: the fall of AICAR with
rising one-carbon supply is enforced by the PPAT--NUDT5 glue, and would not occur
--- indeed would reverse --- in its absence. The cell thus reads its one-carbon and
purine-synthetic status, via the AICAR level set by the glue, into the activity
of a kinase that governs glycolysis, fatty-acid oxidation and biosynthetic
commitment. A NUDT5-null cell, on this account, is predicted to mis-report its
one-carbon status to AMPK, sustaining an AICAR/ZMP tone --- and hence an AMPK tone
--- that is inappropriately high when one-carbon units are plentiful. This connects
a metabolite-glue, a folate-pathway input and an energy-signalling output in a
single testable circuit, and may bear on the metabolic phenotypes of cells with
altered NUDT5 or one-carbon-enzyme expression, including many tumours.

\subsection{Corollaries for thiopurine and antifolate chemotherapy}
\label{sec:chemo}

The PPAT--NUDT5 node is not only a physiological sensor but a pharmacological
target, and the model yields several corollaries for drugs that act on it.

\emph{Thiopurines.} \citet{witus2026} show that thiopurine metabolites
(6-methyl-thio-IMP and related species) are themselves glues of PPAT--NUDT5,
occupying the same pocket as AMP but with distinct geometry and enhanced potency.
In the model such a compound raises $\theta$ towards saturation independently of
the endogenous AMP and PRPP levels. Two consequences follow. First, a
thiopurine-driven increase in $\theta$ shuts PPAT and, by Eq.~\eqref{eq:aicar},
should \emph{lower} the AICAR overshoot rather than raise it --- distinguishing the
glue arm of thiopurine action from the classical antimetabolite arm (incorporation
of thioguanine nucleotides into DNA), and predicting that AICAR is not a marker of
this component of the response. Second, because the glue and PRPP compete, the
efficacy of a thiopurine glue should depend on the cellular PRPP level, and hence
on one-carbon and folate status: conditions that raise PRPP (one-carbon
limitation) should antagonise glue-mediated PPAT inhibition, while conditions that
lower PRPP should potentiate it.

\emph{Antifolates.} Methotrexate and related antifolates deplete
formyl-tetrahydrofolate. In the model this is a move to low one-carbon
availability: ATIC and GART stall, PPAT continues from an ample PRPP pool, and
AICAR accumulates --- the well-documented ``AICAR transformylase'' block. The model
adds that this same low-one-carbon state maximally dissociates the PPAT--NUDT5
glue (high PRPP, low AMP), consistent with the report of \citet{witus2026} that
methotrexate dissociates the endogenous complex. Antifolates and thiopurine glues
thus push the node in \emph{opposite} directions --- one dissociating, one
stabilising the inhibitory complex --- which predicts antagonism if combined, and
identifies PRPP and AICAR as candidate pharmacodynamic markers of where a given
cell sits on this axis.

\emph{Resistance.} \citet{strefeler2025} report that NUDT5 loss confers resistance
to nucleobase analogues that require PRPP-dependent activation (6-thioguanine,
5-fluorouracil) but not to their nucleoside forms, because unrestrained PPAT
depletes the PRPP needed to activate the drug. The model recovers the mechanistic
basis --- the PRPP draw-down of \S\ref{sec:ko} --- and predicts a corollary: the
same NUDT5-null state, by inverting the AICAR coupling, should also blunt any
therapeutic strategy that relies on an AICAR/AMPK response to one-carbon
restriction. Resistance to nucleobase analogues and altered AICAR/AMPK signalling
are, on this view, two faces of the same lesion, and NUDT5 status is predicted to
stratify both.

\subsection{Limitations}

The model is minimal by design. Its free AMP is an output of the adenylate kinase
equilibrium and is structurally low ($2$--$13~\mu$M), which places the AICAR
prediction in the PRPP-gated, energy-charge-silent regime (\S\ref{sec:dualsensor}).
This is a scoping condition rather than an unconstrained assumption: free AMP is
fixed by the equilibrium and the high energy charge of proliferating cells, and
the AMP arm engages only once the energy charge falls, so the prediction is
stated for the high-energy-charge state and would need re-evaluation under energy
stress. The relevant test is the adenylate energy charge, obtainable from
existing nucleotide data, bearing in mind that metabolomics reports total rather
than free AMP. Nine of the thirteen added parameters are assumed, and the
absolute AICAR level is consequently unconstrained (\S\ref{sec:sens}); only its
sign and shape are predicted. The glue also accepts IMP, GMP and AICAR itself as ligands
\citep{witus2026}, which we do not model --- an AICAR-sensitive glue would be
self-reinforcing at low one-carbon supply and could sharpen the predicted
excursion. Finally, PPAT--NUDT5 controls purine synthesis; neither source reports
a route from the complex to growth signalling, so the proliferation law is
retained unchanged from \citet{oizel2020}.

\section{Conclusion}

Incorporating the measured PPAT--NUDT5 metabolite glue into the formate switch
model, with all glue constants fixed by published binding data, yields a single
sharp prediction: AICAR is coupled inversely to one-carbon availability, and this
inverse coupling is enforced by the glue and should reverse its sign in a NUDT5
knockout. The prediction reconciles existing but previously unexplained AICAR
measurements, is independently consistent with the PRPP phenotype of NUDT5-null
cells reported by a second laboratory, and is falsifiable by a direct experiment
that distinguishes the glue from a generic throttle. Its corollaries reach into
AMPK-mediated energy signalling and into the pharmacology of thiopurines and
antifolates, whose actions converge on the same node. We offer the prediction as
a theoretical target and invite its experimental test.

\appendix
\section{Model equations, parameters and numerics}
\label{app:model}

\textbf{State and balances.} The state is
$([\mathrm{AMP}],[\mathrm{ADP}],[\mathrm{ATP}],[\mathrm{For}],[\mathrm{CHO}],
[\mathrm{PRPP}],[\mathrm{GAR}],[\mathrm{AICAR}])$, with
$[\mathrm{THF}]=F_{\mathrm{tot}}-[\mathrm{CHO}]$ and
$c=[\mathrm{CHO}]/(K_c+[\mathrm{CHO}])$. Reversible FTHFS is
\begin{equation*}
v_{\mathrm{FTHFS}}=\frac{h}{F_{\mathrm{tot}}(H+[\mathrm{For}])}
\Big([\mathrm{For}][\mathrm{THF}]
-\tfrac{[\mathrm{CHO}][\mathrm{ADP}][\mathrm{P}_i]}{K_{eq}[\mathrm{ATP}]}\Big),
\end{equation*}
reducing to the published irreversible law as $K_{eq}\to\infty$. The eight
steady-state balances are: formate,
$hf_{SCF}+k_F[\mathrm{For}_x]=v_{\mathrm{FTHFS}}+k_F[\mathrm{For}]$; CHO-THF,
$v_{\mathrm{FTHFS}}+hf_{S,\mathrm{CHO}}=v_{\mathrm{GART}}+v_{\mathrm{ATIC}}
+v_{\mathrm{TYMS}}$; PRPP,
$S_{\mathrm{PRPP}}=v_{\mathrm{PPAT}}+(k_o+\mu)[\mathrm{PRPP}]$; GAR,
$v_{\mathrm{PPAT}}=v_{\mathrm{GART}}+(\mu+k_{\mathrm{deg}})[\mathrm{GAR}]$; AICAR,
$v_{\mathrm{GART}}=v_{\mathrm{ATIC}}+(\mu+k_{\mathrm{deg}})[\mathrm{AICAR}]$;
adenine, $\phi v_{\mathrm{ATIC}}=\mu([\mathrm{AMP}]+[\mathrm{ADP}]
+[\mathrm{ATP}]+\Pi_{\mathrm{ade}})$; energy,
$a\frac{[\mathrm{ATP}]}{A+[\mathrm{ATP}]}
=e_g\frac{[\mathrm{ADP}]}{E_g+[\mathrm{ADP}]}
+e_o\frac{[\mathrm{ADP}]}{E_o+[\mathrm{ADP}]}$; and adenylate kinase,
$[\mathrm{AMP}][\mathrm{ATP}]=K[\mathrm{ADP}]^2$. Here
$v_{\mathrm{TYMS}}=\mu[\mathrm{DNA}]/4$ and $\mu=(-m+a[\mathrm{ATP}]/(A+[\mathrm{ATP}]))/\epsilon$.

\textbf{Numerics.} Concentrations are solved in logarithmic coordinates (CHO-THF
through a logistic map bounded by $F_{\mathrm{tot}}$), enforcing positivity, by
trust-region least squares with restarts from perturbed initial guesses. Each
condition is swept forwards and backwards with continuation; agreement to
$<4\times10^{-14}$ excludes hysteresis. All reported points converge with
residual $<10^{-13}$.

\begin{table}[t]
\centering
\caption{Parameters added to the model of \citet{oizel2020}. Shared parameters
retain their published values. Provenance: fitted to \citet{witus2026} (F),
literature (L), or assumed (A).}
\label{tab:par}
\begin{tabular}{llc}
\toprule
Symbol & Value & Prov. \\
\midrule
$K_{dP}$ & $104~\mu$M (PRPP dissociation) & F \\
$K_a$ & $56.4~\mu$M (AMP glue affinity) & F \\
$n$ & $1.33$ (Hill coefficient) & F \\
$F_{\mathrm{tot}}$ & $0.02$~mM (folate pool) & A \\
$K_c$ & $0.002$~mM (GART/ATIC $K_m$, CHO-THF) & A \\
$K_g,K_t$ & $0.01$~mM (GART/ATIC $K_m$, GAR/AICAR) & A \\
$g_{\max},t_{\max}$ & $5\times$ demand (GART/ATIC capacity) & A \\
$K_{\mathrm{PRPP}}$ & $0.3$~mM (PPAT $K_m$, PRPP) & A \\
$P_{\max}$ & $2\times$ demand (PPAT capacity) & A \\
$S_{\mathrm{PRPP}}$ & $2\times$ demand (PRPP supply) & A \\
$k_o$ & $2~\mathrm{h}^{-1}$ (PRPP drain) & A \\
$k_{\mathrm{deg}}$ & $0.1~\mathrm{h}^{-1}$ (GAR/AICAR drain) & A \\
$K_{eq}$ & $40$ (FTHFS equilibrium) & L \\
$[\mathrm{P}_i]$ & $5$~mM (orthophosphate) & A \\
\bottomrule
\end{tabular}
\end{table}

\section*{Data availability}
This is a theoretical study and generates no new experimental data. The published
data used for calibration and comparison are available in the cited sources:
\citet{witus2026}, \citet{strefeler2025} and \citet{oizel2020}. The model outputs
that underlie the figures are provided in the code repository below.

\section*{Code availability}
The code that implements the model, performs the calibration and produces the
figures is available at \url{https://github.com/av2atgh/Formate-NUDT5}. Every parameter
carries an inline provenance tag (fitted, literature, or assumed).

\section*{Acknowledgements}
Nodes \& Links Ltd provided support in the form of salary for Alexei Vazquez, but
did not have any additional role in the conceptualization of the study, analysis,
decision to publish, or preparation of the manuscript. The code and text were
prepared with the assistance of Claude Opus~4.8, an AI assistant developed by
Anthropic. A.V.\ directed the use of the AI assistant, verified all results, and
takes full responsibility for the content.

\section*{Competing interests}
The author declares no competing interests.

\bibliographystyle{cas-model2-names}

\end{document}